\documentclass[11pt]{article}
\usepackage{fleqn,cospar}

\usepackage{url}

\usepackage{graphicx}
\usepackage[figuresright]{rotating}

\title{X-RAY BURSTS FROM THE GALACTIC X-RAY TRANSIENT SOURCE  GRS 1915+105}

\author{J. S. Yadav\address{Tata Institute of Fundamental Research, Homi Bhabha Road, Mumbai,
400 005, India}, and  A. R. Rao$^{1}$}  

\begin{document}

\maketitle

\begin{abstract}
 We have analyzed publicly available   RXTE/PCAs  archival data
of GRS 1915+105 during its burst/flaring state. The burst cycle
ranges from 30 to 1300 s. These bursts are different from the type I 
and type II classical bursts seen in Low Mass X-ray Binaries (LMXBs) in
terms of their temporal and spectral properties. We have classified 
these bursts on the basis of properties observed during the quiescent
(low flux) phase. The 2 - 10 Hz QPOs are present during the quiescent     
phase and disappear during the burst phase of all types of these 
X-ray bursts. The duration of the quiescent phase can be explained 
assuming an outflow from the post-shock regions and the catastrophic
Compton cooling.

\end{abstract}

\section*{INTRODUCTION}
The Galactic X-ray transient source GRS~1915+105 has shown spectacular 
X-ray variability during last four years of its observation by RXTE and other 
satellites (Greiner et al. 1996, Yadav et al. 1999, Belloni et al. 2000). 
This source  was 
discovered in 1992 with the WATCH 
all sky X-ray monitor onboard the GRANAT satellite (Castro-Tirado et al. 1994).
 The X-ray intensity was found to vary on a variety of time scales and
the light curve showed a complicated pattern of dips and rapid transitions 
between high and low intensity (Belloni et al. 1997, 
Taam et al. 1997).
 Recently, Yadav
et al. (1999)  have made a detailed study of various types of X-ray bursts
seen in GRS~1915+105 from IXAE/PPCs observations during 1997 June - August 
and  have suggested  that during the bursts the source switches back and 
forth between the 
low-hard state and the high-soft state  near  critical accretion rates in 
a very short time scale. The fast time scale for the transition of the 
state is explained by invoking the appearance and disappearance of the 
advective disk in its viscous time scale.

\begin{table*}
\begin{center}
\centerline{Table I.}
\centerline{Summary of selected observations of GRS~1915+105} 
\begin{tabular}{llclcccl}
\hline\hline     
Observation ID & Date &   Exposure & Type of  &ASM$^a$ & Rec.&Av. Q.& Remark\\ 
 & (UT) &   (s) & Bursts & Flux &Time (s)&Flux&\\
\hline  
 10408-01-01-01& 1996 April 06 05:40& 5600 &Regular (R2)&99.4&280$^{b}$&6200&QT $\sim$ 20 s\\ 
 10408-01-38-00& 1996 Oct 07 05:44 & 7000 &Regular (R3)&98.8&1150$^{b}$&3200&-\\ 
 20402-01-30-00& 1997 May 26 12:25 & 3215 &Regular (R1b)&47.6&105$^{b}$&7500&BT $\sim$ 20 s\\ 
 20402-01-33-00& 1997 June 18 14:17 & 3472 &Irregular (IR)&61.5& var.& var.&- \\ 
 20402-01-34-01& 1997 June 22 19:27 & 2550 &Regular (R1a)&59.7&55$^{b}$&8700&BT $\sim$ 20 s\\ 
\hline  
\end{tabular}
\end{center} 
{QT = quiescent time, BT = burst time, Av. Q. Flux= average quiescent time flux (c/s), var. = variable} \hspace{4mm} 
$^{a}$ {Mean ASM flux (c/s) for a day},  
$^{b}$ {Mean burst recurrence time}
\end{table*}

In this paper, we  present  results of our   analysis of 
a set of publicly available  RXTE/PCA 
observations of   GRS 1915+105 during last 
four years which samples a broad range of burst cycles  from 30 s to 1300 s.  
Each burst cycle consists of a low flux quiescent phase
followed by a high flux burst phase and the fast transition in less than 10 s.
We include here only those bursts which have dips or quiescent phase with
hard spectrum (Yadav 2001, Yadav \& Rao 2001). 
\section*{OBSERVATIONS}
\vspace{2mm}
\begin{minipage}{85mm}
\hspace{5mm} We have selected a set of observations from publicly available RXTE/PCA data
for the  X-ray transient source GRS 1915+105 (Jahoda et al. 1996). The source was in high/flaring state
during these observations. The dips or the quiescent phase in 
these observations are spectrally hard while the brighter portions (the
burst phase) are soft.
A portion of  2$-$13 keV light curves
for different days added for all PCA units (except on 1996 April 6, when only
three PCA units were on), are shown in Figure 1.
 In Table 1 we list details of these observations alongwith some of the
properties of the observed X-ray bursts. During these observations ASM flux 
varies from 47 to 100 ASM c/s while the quiescent time flux varies in the
range of 3000$-$10000 c/s (added for 5 PCA units) in the 2$-$13 keV 
energy range. 
\section*{RESULTS AND DISCUSSION}
\vspace{2mm}
\hspace{5mm} These bursts can broadly be put into two classes: regular bursts centered 
around a fixed period with  low dispersion ($\delta P / P \sim 1 - 50 \%$)
and irregular bursts with no fixed periodicity ($\delta P / P >  50 \%$).
The observed bursts are classified into four 
types:  (1) regular bursts (R1)  with short burst phase
 lasting for $\sim$ 20 s and recurring every 50$-$145 s (extreme ends of 
this type are shown  in first  and second 
panels from top of Figure 1 (R1a type and R1b type respectively), for other bursts of this type see Vilhu \&
Nevalainen (1998)). These bursts have lowest  dispersion 
($\delta P / P \sim 1 - 10 \%$), 
(2) regular bursts (R2) with short quiescent phase lasting for 
\end{minipage}
\hfil\hspace{\fill}
\begin{minipage}{85mm}
\includegraphics[width=85mm]{jsyadav_fig1.ps}

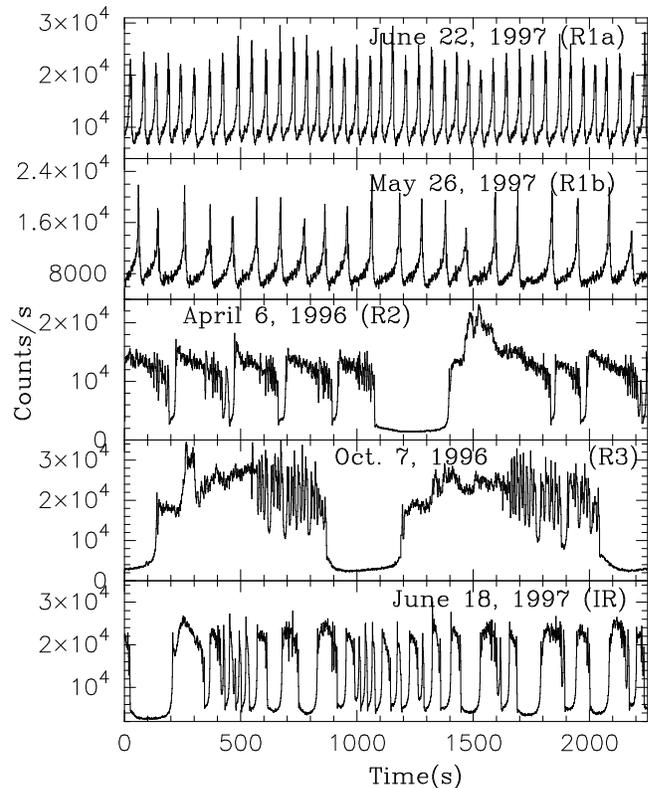
{\sf Fig. 1. The  regular bursts with $\sim$ 55 s recurrence time
(first panel from the top),  regular bursts with $\sim$ 105 s recurrence time
(second panel), regular bursts with $\sim$ 280 s (third panel), regular
bursts with $\sim$ 1150 s recurrence time (fourth panel), and irregular
bursts (fifth panel) observed in GRS~1915+105 with all the PCAs except on
1996 April 6 when only three PCA units were on.}
\end{minipage}
 $\sim$ 20 s
 recurring every 280 s (third panel from top). The $\delta P / P$ is 
upto 50\% for these bursts, 
(3) long regular bursts (R3) recurring every 1150 s (fourth panel from top).
  The $\delta P / P$ is upto 15\% for these bursts,  
 and (4) irregular  bursts (IR) with duration of a few tens to a few hundred
seconds (bottom panel of Figure 1).  The regular bursts with short
burst phase usually have two-peak structure while irregular  and 
long regular bursts show multi-peak structure with higher variability near 
the end of the burst.

The most striking features of these  bursts  are 
slow exponential rise, sharp linear decay and hardening of spectrum as
burst progresses (Paul et al. 1998). The decay time scales are shorter than 
the rise time scales.
In sharp contrast, the decay time is longer than the rise time  in 
classical bursts and  spectrum is initially hard and becomes softer as
the burst decays.  
 The ratio of   luminosity in   type I X-ray bursts ($L_b$) and
the average quiescent X-ray luminosity ($L_q$) is
$ {L_b\over L_q} \sim 10^{-2}$. The time-averaged
type II burst luminosity is  higher, usually 0.4 to 2.2 times the
average luminosity of quiescent emission (Lewin et al. 1995).
The time-averaged luminosity of the  bursts seen in  
GRS~1915+105 is found to be from  0.15 to 4.5 times the average luminosity of 
the  quiescent
emission.  The emission process involved in producing these  bursts is 
likely to be   gravitational as in the case of type II bursts due to the 
energetics involved, but the large value of this ratio indicates the black hole
nature of the compact object where more gravitational potential energy is
available.

Yadav et al.  (1999)   have  suggested that the source is 
in a high-soft state during the 
burst phase and in a low-hard state during the quiescent phase  on the basis of available spectral observations and derived disk parameters of 
GRS~1915+105.
The source 
makes state transitions in very short time scales corresponding to the 
rise and fall time of the bursts (a few seconds). Such fast changes of 
states are possible in the Two Component Accretion Flows (TCAF) where the 
advective disk covers the standard thin disk (Chakrabarti 1996, 
Rao et al. 2000).  
The time scales of the burst phase are  compared  with the viscous time 
scales of the thin accretion disk.  
Recently, Chakrabarti (1999) have presented a solution to the rapid
state transition based on TCAF (Chakrabarti \& Titarchuk 1995). 
The mass outflow from the
regions of the shock compressed flow initiates the quiescent phase and the
catastrophic Compton cooling of the material in the sonic sphere marks the
end of the quiescent phase and the start of the burst phase. This model 
essentially reinforces the suggestion of Yadav et al. (1999) but  gives
a physical basis  for the start of the event.  
The shock compressed gas with compression ratio R $>$ 1
will produce the outflows which pass through the sonic points at 
 R$_c = f_o$ $\times$ R$_s$  provided  the flow is isothermal
till R$_c$, where R$_s$ is the shock location and 
$f_o =$ R$^2$/(R-1)  (Chakrabarti 1999). 
Chakrabarti \& Manickam (2000) have expanded this model further and derived
a correlation between t$_{off}$  the duration of 
off state (duration in which the sonic sphere becomes ready  for
catastrophic Compton cooling) 
and  $\nu_I$ the QPO frequency between 2 $-$
10 Hz. For an average shock 2.5 $<$ R $>$ 3.3, t$_{off}$ is insensitive to the 
compression ratio. Using average value of R $=$ 2.9 and a constant velocity
post-shock flow $\alpha =$ 1 following relation is derived between t$_{off}$ 
and $\nu_I$;  
\begin{equation}
t_{off} = 461.5 \left( {0.1 \over {\Theta_{\dot{M}}}} \right) \left( {m \over
{10}} \right)^{-1} \left( {v_o \over {0.066}} \right)^2 \nu_I^{-2}  s. 
\end{equation}
\begin{minipage}{85mm}
where m is the mass of the black hole in units of the solar mass m $=$ 10,
$v_o =$ 0.066, and $\Theta_{\dot{M}}$ is a dimensionless 
parameter defined as
$\Theta_{\dot{M}} = (\Theta_{out} / \Theta_{in}) \times \dot{m}_d$ where
$\Theta_{in}$ and $\Theta_{out}$  are the solid angles  of the inflow  \& 
outflow respectively and $\dot{m}_d$ is the disk accretion rate in units 
of Eddington accretion rate. In this configuration $\dot{m}_h$ $\sim$ 1 and 
$\dot{m}_d \sim$ 0.1 keeping the total accretion rate $\dot{m}_t$  
close to 1. 

\hspace{5mm} In Figure 2, we plot Eq. 1 in the log - log scale taking t$_{off}$ 
as the quiescent time for $\Theta_{\dot{M}}$ $=$ 0.0145, 0.0245 and 0.0335.
We searched for QPO peaks in the PDS by fitting frequency intervals between 
0.5$-$10 Hz with Gaussian profiles on top of a power-law background 
continuum. The width and position of the Gaussian are kept as free parameters.
Results are plotted in Figure 2. 
The data of bursts with quiescent time in a narrow range are
clubbed together to improve the statistics of PDS analysis. 
The errors in QPO frequency is less than the size of the symbols.
Although we
have used almost similar number of bursts for R1 (a \& b), IR and R3 bursts
as used by Chakrabarti \& Manickam (2000), the data points are reduced in 
our case. We have added the R2 bursts to this analysis which have enabled
us to investigate the inverse-square law dependence over a range of
25$-$320 s (dotted line). In comparison, Chakrabarti \& Manickam (2000)
have few points around 320 s.  
 These results are in good  agreement. The data points 
\end{minipage}
\hfil\hspace{\fill}
\begin{minipage}{85mm}
\includegraphics[width=85mm]{jsyadav_fig2.ps}

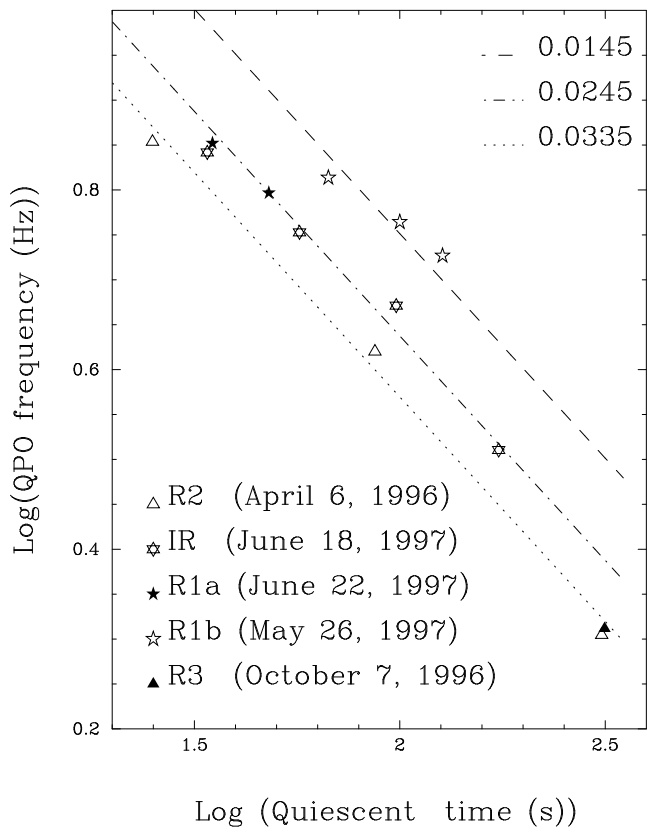
{\sf Fig. 2.  Variation of QPO frequency $\nu_I$ (minimum) with the quiescent time for different types of X-ray bursts observed in GRS~1915+105 (data
points). Plotted lines are the quiescent time $\alpha$ $\nu_{I}^{-2}$ for
different values of ${\Theta}_{\dot{M}}$ (for further details see in text). }
\end{minipage}
\vspace{3mm}

\noindent  
of R2 and R3 bursts when ASM flux was 99.4 \& 98.8 c/s respectively fall
along the dotted line with $\Theta_{\dot{M}} =$ 0.0335. The data points
of R1a and IR bursts when ASM flux was 59.7 and 61.5 c/s respectively
fall along the dashed -  dotted line ($\Theta_{\dot{M}} =$ 0.0245). The 
data points of R1b bursts  during which ASM flux has lowest value of 47.6 c/s
fall along the dashed line ($\Theta_{\dot{M}} =$ 0.0145). It may be 
noted here that the $\Theta_{\dot{M}}$ and the ASM flux though determined
independently agree well for different types of bursts  as both of these
are related to the disk accretion rate $\dot{m}_d$. 

Our results in Figure 2 suggest that the t$_{off}$ represents the quiescent 
time of all the bursts which may or
may not be of the order of the viscous time scales of the thin accretion
disk. However non-zero burst time would represent the viscous time scales
of the thin accretion disk as any change in the $\dot{m}_d$ would require the 
viscous time scale to reach R$_c$  when system can revert back
to the quiescent phase (Yadav et al. 1999).
The IR bursts are produced due to viscous - thermal - instability
and  the quiescent  and  burst time are correlated  suggesting that 
both these parameters  represent the  
 viscous time scales of the thin disk (t$_{off}$ is of the order of viscous
time scales and changing continuously).   The burst time varies over a large 
range of 100 to 500 s
suggesting viscous - thermal condition are not stable during R2 bursts. 
The dispersion around a fixed period is large ($\delta P / P$ is upto  50 \%).
The burst duration of R3 bursts fall in a range of 700 to 1000 s suggesting 
fairly stable viscous - thermal conditions during R3 bursts. 
The quiescent time is fixed during R2 and R3 bursts ($\sim$ 20 \& $\sim$ 320 s 
respectively) which represents t$_{off}$.

 The R1 bursts seen on 1997 May 26 and 1997 June 22 termed as ``ringing
flares'' have a short burst phase of $\sim$ 20 s.  The peak flux during the 
burst phase lasts only for 1 - 2 s . The quiescent time 
properties of these bursts are very different from the properties observed for 
other types of bursts. The average quiescent flux is high (7000$-$10000 c/s) 
and the $\Gamma$ of the energy spectrum is 2.67.  The ASM flux is low 
(47$-$60 ASM c/s). The HR$_1$  (ratio of flux in 5$-$13 keV band and flux in 
2$-$5 keV band) is high (1.0$-$1.06) during the 
quiescent phase.  The ratio of the average burst flux and  the average 
quiescent flux is almost constant during these bursts (Yadav \& Rao 2001).
The  dispersion around a fixed period is lowest ($\delta P / P \sim 1 - 10 \%$).
This type of bursts have been observed for extended period almost
continuously from 1997 May 26 to June 26 suggesting very stable 
thermal - viscous conditions during these burst (Yadav et al. 1999, 
Yadav \& Rao 2001). 
At t$_{off}$, the 
sonic sphere cools down which ends the quiescent phase and marks the start
of the burst phase.
 However the conditions are very stable and  remain
unchanged. The outflows immediately start and sub-Keplerian halo appears 
quickly which abruptly ends the burst
phase and the next quiescent phase starts producing ringing type burst phase  
with the burst peak flux hardly lasting for  1 - 2 s.  The quiescent time
decreases   
as the $\dot{m}_d$ increases (Figure 2). As the average quiescent flux 
increases the t$_{off}$ decreases reducing the burst recurrence time 
without affecting the burst phase duration.  An  
increase in the $\dot{m}_d$  increases  the quiescent flux
and reduces the burst cycle (appearing of the burst phase more frequently) 
which explains why this
type of bursts were observed over a large range of ASM flux from 47 to 60 c/s
 (the HR$_1$   remains unchanged).

\end{document}